\documentclass[sigconf]{acmart}
\AtBeginDocument{%
  }

\copyrightyear{2026}
\acmYear{2026}
\setcopyright{cc}
\setcctype{by}
\acmConference[UIST '26]{The 39th Annual ACM Symposium on User Interface Software and Technology}{November 02--05, 2026}{Detroit, MI, USA}
\acmBooktitle{The 39th Annual ACM Symposium on User Interface Software and Technology (UIST '26), November 02--05, 2026, Detroit, MI, USA}
\acmDOI{10.1145/3830398.3830585}
\acmISBN{979-8-4007-2856-3/2026/11}

\usepackage{xcolor}
\usepackage{listings}
\definecolor{jsontoplevel}{HTML}{1596C8} 
\definecolor{jsonsection}{HTML}{2AA6D2}  
\definecolor{jsonfield}{HTML}{9A7BD1}    
\definecolor{jsonstring}{HTML}{E66A2C}   
\definecolor{jsonnumber}{HTML}{66A61E}   
\definecolor{jsonpunct}{HTML}{888888}    
\definecolor{jsonliteral}{HTML}{8E63CE}  

\lstdefinelanguage{prettyjson}{
  morestring=[b]",
  stringstyle=\color{jsonstring},
  showstringspaces=false,
  breaklines=true,
  breakatwhitespace=false,
  keepspaces=true,
  columns=fullflexible,
  literate=
   {"SMILES"}{{{\color{jsontoplevel}"SMILES"}}}{8}
   {"Goal"}{{{\color{jsontoplevel}"Goal"}}}{6}
   {"intent"}{{{\color{jsontoplevel}"intent"}}}{8}
   {"property_constraints"}{{{\color{jsonsection}"property\_constraints"}}}{22}
   {"structure_constraints"}{{{\color{jsonsection}"structure\_constraints"}}}{23}
   {"edit_plan"}{{{\color{jsonsection}"edit\_plan"}}}{11}
   {"edit_groups"}{{{\color{jsonsection}"edit\_groups"}}}{13}
   {"property_key"}{{{\color{jsonfield}"property\_key"}}}{14}
   {"current_value"}{{{\color{jsonfield}"current\_value"}}}{15}
   {"target_value"}{{{\color{jsonfield}"target\_value"}}}{14}
   {"molecule"}{{{\color{jsonfield}"molecule"}}}{10}
   {"anchor_ids"}{{{\color{jsonfield}"anchor\_ids"}}}{12}
   {"avoid_ids"}{{{\color{jsonfield}"avoid\_ids"}}}{11}
   {"free_form"}{{{\color{jsonfield}"free\_form"}}}{11}
   {"pos"}{{{\color{jsonfield}"pos"}}}{5}
   {"prompt"}{{{\color{jsonfield}"prompt"}}}{8}
   {"operation"}{{{\color{jsonfield}"operation"}}}{11}
   {"target"}{{{\color{jsonfield}"target"}}}{8}
   {true}{{{\color{jsonliteral}true}}}{4}
   {false}{{{\color{jsonliteral}false}}}{5}
   {null}{{{\color{jsonliteral}null}}}{4}
   {0}{{{\color{jsonnumber}0}}}{1}
   {1}{{{\color{jsonnumber}1}}}{1}
   {2}{{{\color{jsonnumber}2}}}{1}
   {3}{{{\color{jsonnumber}3}}}{1}
   {4}{{{\color{jsonnumber}4}}}{1}
   {5}{{{\color{jsonnumber}5}}}{1}
   {6}{{{\color{jsonnumber}6}}}{1}
   {7}{{{\color{jsonnumber}7}}}{1}
   {8}{{{\color{jsonnumber}8}}}{1}
   {9}{{{\color{jsonnumber}9}}}{1}
   {:}{{{\color{jsonpunct}:}}}{1}
   {,}{{{\color{jsonpunct},}}}{1}
   {\{}{{{\color{jsonpunct}{\{}}}}{1}
   {\}}{{{\color{jsonpunct}{\}}}}}{1}
   {[}{{{\color{jsonpunct}[}}}{1}
   {]}{{{\color{jsonpunct}]}}}{1}
}

\lstdefinestyle{compactjson}{
  language=prettyjson,
  basicstyle=\ttfamily\fontsize{7.5pt}{7.2pt}\selectfont,
  backgroundcolor=\color{white},
  frame=none,
  numbers=none,
  showstringspaces=false,
  columns=fullflexible,
  keepspaces=true,
  breaklines=true,
  breakatwhitespace=false,
  tabsize=2,
  xleftmargin=0.15em,
  xrightmargin=0.15em,
  aboveskip=0pt,
  belowskip=0pt,
  lineskip=-0.15pt
}

\usepackage{xspace}
\usepackage{enumitem}
\begin{document}

\newcommand{\tool}{\textit{MolecularCanvas}\xspace}
\newcommand{\revise}[1]{{#1}}

\title{\tool: LLM-assisted Small-Molecule Drug Discovery via Structure-Guided Constraints}

\author{Haoyu Dong}
\authornote{Both authors contributed equally to this research.}
\affiliation{%
  \institution{Zhongnan University of Economics and Law}
  \city{Wuhan}
  \country{China}}
\email{dan1el131@stu.zuel.edu.cn}
\affiliation{%
  \institution{University of Glasgow}
  \city{Glasgow}
  \country{United Kingdom}}
\email{3156853d@student.gla.ac.uk}

\author{Rui Sheng}
\authornotemark[1]
\affiliation{%
  \institution{The Hong Kong University of Science and Technology}
  \city{Hong Kong}
  \country{China}
}
\email{rshengac@connect.ust.hk}

\author{Shuhao Zhang}
\affiliation{%
  \institution{Carnegie Mellon University}
  \city{Pittsburgh}
  \state{Pennsylvania}
  \country{USA}
}
\email{shuhaozh@alumni.cmu.edu}

\author{Yushi Sun}
\affiliation{%
  \institution{The Hong Kong University of Science and Technology}
  \city{Hong Kong}
  \country{China}
}
\email{ysunbp@connect.ust.hk}

\author{Dingyang Wu}
\affiliation{%
  \institution{The Hong Kong University of Science and Technology}
  \city{Hong Kong}
  \country{China}
}
\email{dwubl@connect.ust.hk}

\author{Hanxiang Chao}
\affiliation{%
  \institution{Wuhan University}
  \city{Wuhan}
  \country{China}
}
\email{chx_whu@whu.edu.cn}

\author{Olexandr Isayev}
\affiliation{%
  \institution{Carnegie Mellon University}
  \city{Pittsburgh}
  \state{Pennsylvania}
  \country{USA}
}
\email{olexandr@olexandrisayev.com}

\author{Huamin Qu}
\affiliation{%
  \institution{The Hong Kong University of Science and Technology}
  \city{Hong Kong}
  \country{China}
}
\email{huamin@cse.ust.hk}

\author{Yuyang Wu}
\authornote{Corresponding authors.}
\affiliation{%
  \institution{Carnegie Mellon University}
  \city{Pittsburgh}
  \state{Pennsylvania}
  \country{USA}
}
\email{yuyangwu@andrew.cmu.edu}

\author{Yanna Lin}
\authornotemark[2]
\affiliation{%
  \institution{University of Waterloo}
  \city{Waterloo}
  \state{Ontario}
  \country{Canada}
}
\email{yanna.lin@uwaterloo.ca}

\renewcommand{\shortauthors}{Dong et al.}
\newcommand{\rui}[1]{\textcolor{orange}{#1}}
\newcommand{\yanna}[1]{\textcolor{teal}{#1}}
\newcommand{\haoyu}[1]{{#1}}
\newcommand{\yuyang}[1]{\textcolor{orange}{#1}}
\newcommand{\ie}{{i.e.,}\xspace}
\newcommand{\eg}{{e.g.,}\xspace}
\newcommand{\ea}{{et~al.}\xspace}
\newcommand{\etal}{{et~al.}\xspace}
\newcommand{\aka}{{a.k.a.}\xspace}
\newcommand{\etc}{{etc\xperiod}\xspace}
\renewcommand{\figurename}{Fig.}
\renewcommand{\tablename}{Tab.}
\renewcommand{\sectionautorefname}{Sec.}
\renewcommand{\subsectionautorefname}{Sec.}
\renewcommand{\subsubsectionautorefname}{Sec.}

\begin{abstract}
  Small-molecule drug discovery relies on iterative molecular optimization, where chemists repeatedly modify candidate compounds to balance multiple competing properties such as efficacy, toxicity, and solubility. Recent advances in generative AI (GenAI), have shown promise in accelerating this process by automatically proposing new molecular structures or targeted modifications.
  However, existing GenAI-based molecular design tools remain poorly aligned with experts’ real-world workflows. 
  Specifically, they offer limited support for specifying structure-level modification intents on molecules, provide insufficient transparency into model-generated modifications, and lack integrated support for downstream property evaluation with external computational tools.
  To address these challenges, we introduce \tool, an interactive system that enables users to iteratively construct an optimization context by integrating high-level goals, structure-level annotations, property constraints, and reference-based preferences. This context guides the generation of candidate molecules across diverse molecular structures.
  \tool further enhances transparency by providing evidence for AI-generated suggestions, and streamlines molecular evaluation by integrating commonly used property computational tools into a unified interface.
  Finally, a user study with 12 participants demonstrates the usefulness and effectiveness of \tool in helping users optimize candidate molecules.
\end{abstract}


\maketitle

\section{Introduction}

Small-molecule drug discovery is critical in therapeutic development, as it enables the identification and optimization of compounds with desirable biological and pharmacological properties \cite{hughes2011principles}.
In practice, researchers often begin with a candidate molecule identified through screening or prior studies. They then iteratively modify the molecule, such as by replacing functional groups or adjusting substructures, and evaluate whether these changes lead to improved properties~\cite{tabana2023target}. 
Through repeated cycles of modification, candidate molecules are gradually refined into viable drug leads.


Molecular optimization is inherently complex and iterative, requiring careful trade-offs among multiple properties, such as toxicity, solubility, and synthesizability \cite{gao2022benchmark}.
To accelerate molecular design, various AI approaches have been developed to predict molecular properties and screen candidate compounds, enabling experts to iteratively modify candidate molecules and test hypotheses based on model predictions \cite{wu2018moleculenet}. More recently, advances in generative AI (GenAI), particularly the emergence of large language models (LLMs), have enabled the automated generation of new molecular structures or targeted modifications to existing molecules, further reducing the manual effort required from domain experts \cite{loeffler2024reinvent}. For instance, recent transformer-based models such as Chemformer \cite{chemformer2022} can generate new candidate molecules or suggest structural modifications by learning patterns from large molecular datasets.

Despite recent progress, current GenAI-based molecular design systems remain far from fully autonomous and still require substantial human involvement in practice \cite{gao2022benchmark,menke2024metis}.
However, chemists face three key challenges to effectively edit molecules with the assistance of GenAI, arising from the limitations of existing tools.
First, researchers face difficulty in externalizing their modification requirements or domain knowledge when modifying generated molecules with the help of GenAI. For example, chemists might aim to preserve key structural components, such as scaffolds or important substructures, while modifying specific regions of a molecule. However, many generative systems rely on natural language–based interfaces with limited mechanisms for specifying such constraints, making it challenging to ensure that AI-generated modifications align with expert intent.
In addition, recommendations from generative models are often presented without clear evidence or provenance, making it difficult for chemists to evaluate or justify the proposed changes in collaborative drug discovery settings~\cite{sun2025kerag,yang2024crag}.
Finally, chemists frequently need to rely on multiple platforms to inspect, edit, and evaluate GenAI-generated molecules.
This is because different platforms often support different parts of the workflow, such as molecule editing, property calculation, or downstream analysis, but rarely provide integrated support.
As a result, researchers must repeatedly switch between platforms to complete a single design iteration, which makes the workflow fragmented, time-consuming, and prone to error. For example, a researcher may use a generative model to generate candidate molecules, a separate cheminformatics toolkit such as RDKit to compute physicochemical properties like LogP and QED, an independent web-based platform such as SwissADME or ADMETlab to evaluate ADMET profiles, and yet another tool to inspect and manually edit molecular structures, requiring repeated context switches across disconnected environments within a single design iteration \cite{loeffler2024reinvent, daina2017swissadme, xiong2021admetlab}.

To address these challenges, we conducted a formative study with six domain experts in medicinal chemistry and computational drug discovery. Through semi-structured interviews, we examined how they currently modify candidate molecules with GenAI and derived five concrete design requirements. Based on these insights, we present \tool, an interactive tool for GenAI-assisted molecular optimization. 
Rather than relying on a single free-form prompt, \tool allows users to iteratively construct an optimization context by integrating high-level goals, structure-level annotations, property constraints, and reference-based preferences. This structured context is then used to guide the generation of candidate molecules across diverse molecular structures.
After each modification, GenAI provides traceable and verifiable evidence, such as links to relevant literature, database records, or prior-art checks. In addition, we integrate computational interfaces from multiple commonly used platforms into \tool, reducing the need to switch between tools and streamlining the evaluation of generated molecules.
In summary, we make three contributions:

\begin{itemize}[leftmargin=10pt]
    \item A formative study with medicinal chemistry experts that reveals interaction challenges and design requirements when collaborating with GenAI for molecular optimization.
    
    \item \tool, a novel interactive system that enables chemists to directly interact with molecular structures, specify design constraints, and iteratively refine AI-generated modifications.
    
    \item A user study with 12 participants demonstrating the usefulness and effectiveness of \tool for supporting AI-assisted molecular design workflows.
\end{itemize}
\section{Related Work}
In this section, we mainly introduce the background of small-molecule design and the related algorithms or systems
to better clarify our research gap.

\subsection{Small-Molecule Optimization}

Small-molecule drug discovery plays a central role in modern therapeutic development~\cite{beck2022small,maurer2022designing,racz2025changing}. 
In practice, experts often need to iteratively modify candidate compounds to improve their overall drug profile,  balancing multiple objectives such as potency, selectivity, and solubility~\cite{hughes2011principles,waring2015analysis,hann2012finding,bickerton2012quantifying}.
Each iteration typically involves adjusting key scaffolds or functional groups, making localized structural changes~\cite{papadatos2010mmp,kenny2011mmp}.
Specifically, chemists must consider not only whether a modification improves target properties, but also whether it preserves essential substructures, remains chemically reasonable, and supports downstream synthesis and validation. 
After that, experts will evaluate the modified results based on computational metrics or real experimental evidence~\cite{sadybekov2023computational,lam2025navigating}. 
Molecular optimization is therefore not simply a search problem in chemical space, but a design activity in which structural constraints, trade-offs, and evaluation context must be maintained across iterations. These characteristics motivate multiple automated algorithms and interactive systems that support fine-grained modification intent~\cite{kingma2013vae,jin2018junctiontree,chemformer2022,edwards2022translation}.

\subsection{AI-Assisted Molecular Design}

In computational molecular design, the generation of candidate structures based on specific requirements has become increasingly efficient and accessible with the advent of generative AI, particularly large language models (LLMs) ~\cite{bran2024augmenting,schwaller2021prediction,ozcelik2025generative,stokes2020deep}.
Prior work has explored various representations of molecules, such as strings and graphs, enabling both de novo generation and local structural modification through approaches including variational autoencoders, graph-based generative models, and transformer-based chemistry models~\cite{kingma2013vae,jin2018junctiontree,chemformer2022,edwards2022translation}. 
Building on these representations, many of these methods focus on goal-directed optimization, where molecules are iteratively refined to satisfy desired properties or objectives~\cite{loeffler2024reinvent,jensen2019graphga,kim2024gflownet}. 

However, considering the imperfection of GenAI, several interactive systems have been proposed to help experts work with GenAI to design molecules. 
In these systems, human experts are not merely recipients of AI-generated outputs, but active participants in specifying objectives, steering exploration, and evaluating candidate molecules.
For example, \revise{ChatChemTS provides an LLM-powered conversational chatbot that allows chemists to specify molecular design objectives~\cite{ishida2025chatchemts}
}.
ChemCrow~\cite{bran2024augmenting} demonstrates an LLM-centered chemistry workflow in which users interact with the system through natural-language instructions and queries.
Other works can support navigation in latent molecular neighborhoods, allowing users to guide candidate search through property or substructure constraints~\cite{zhang2024chemnav,zheng2023desirable}. 

Despite this progress, most existing systems support only limited forms of user interaction \revise{for accessing or configuring molecular design tools}, relying primarily on natural language. This significantly hinders users’ ability to work with molecular structures, as even simple queries about a molecule become cumbersome to express and interpret.
For example, chemists might need to communicate finer-grained design intent, including permissible editing scope, structures to preserve, and the evidential basis for evaluating proposed changes~\cite{zhang2024deeplead,lam2025navigating}.
Therefore, our work aims to support a more expressive human-in-the-loop molecular design approach with GenAI, enabling experts to specify not only desired outcomes but also finer-grained structural and evidential requirements.
\subsection{In-Context Prompting Tools for GenAI}
Recent progress in generative AI has renewed interest in interaction paradigms that make users steer model behavior through prompts~\cite{arawjo2024chainforge,jiang2022prompting,Zhu2024promptbench,wu2022aichains,sheng2023knowledge}. Prior works show that such approaches can provide flexible control.
However, users often struggle to formulate effective prompts across iterative tasks, especially when prompts must capture multimodal, spatial, or structured information that cannot be easily specified through text alone~\cite{brown2020language,liu2023pretrainpromptpredict,zamfirescu2023whycantprompt,wu2022aichains}.
This problem motivates growing research exploring in-context prompting beyond plain text alone.
Some works extend in-context prompting through persistent prompt-engineering structures, such as chained prompt workflows \cite{arawjo2024chainforge, song2026vizdefender} \revise{and modular, composable prompt components~\cite{kim2023cells}}. 
Other works \revise{extend prompting through direct  manipulation or} visual and spatial context, allowing users to instruct generative models through \revise{operations on generated objects,} multimodal prompt composition, reference images, or canvas-based interactions \cite{masson2024directgpt, peng2024designprompt,choi2024creativeconnect,chung2023promptpaint, lin2023inksight, song2025gvvst}.

\revise{
Our work builds on this line of research in the context of molecular optimization, where users need to communicate complex and interdependent modification requirements.
Existing systems, however, are not designed to support specific requirements of molecular optimization, such as localized structural edits, property goals, anchor and avoidance constraints, and preferences derived from reference molecules.
\tool{} addresses this gap by providing structure-grounded interactions for expressing these requirements and integrating them with a validation-aware generation pipeline that produces chemically valid candidate molecules.
}

\section{Formative Study}
\label{sec:formative_study}
We conducted a formative study with six medicinal chemists to investigate challenges in generative AI–assisted molecular design workflows and to inform the design of human–AI collaborative support tools for molecular design.

\subsection{Setup}

\subsubsection{Participants.} 
We recruited six participants (1 female and 5 males; mean age $32.83 \pm 8.80$, identified as E1--E6), with professional experience in small-molecule drug discovery and computational chemistry. Participants were recruited through academic referrals, targeted outreach within medicinal chemistry communities, and social media advertisements.
All participants have hands-on experience in small-molecule drug design and prior exposure to generative AI tools for molecular generation or optimization.
The participant pool included both academic and industry professionals: 4 PhD students, an industry computational chemist, and a principal investigator leading a translational drug discovery program.
Their experience in drug discovery ranged from approximately 2 to over 10 years, with an average of 5.8 years.

\subsubsection{Procedure.} 
We conducted one-on-one semi-structured online interviews, each lasting approximately 30 minutes. 
All participants provided informed consent prior to the interview and agreed to audio recording for research purposes.

The interviews began with background questions about participants’ demographic information, professional roles, experience in small-molecule drug discovery, and prior use of AI-assisted molecular design tools.
Participants were then asked to describe their typical workflows for molecular optimization after virtual screening, including how they modify, evaluate, and iterate on candidate molecules.
We encouraged participants to walk through recent design cases to provide additional contexts and concrete details.
To identify workflow challenges, we asked open-ended questions such as ``What difficulties do you encounter when optimizing candidate molecules?'', ``What challenges arise during optimization?'', and ``What limitations do you experience when using AI-based or LLM-based tools for molecular design?''. 
We further used follow-up questions to encourage participants to elaborate on specific situations and examples, such as ``Can you explain in more detail why this aspect is challenging?'' and ``Can you provide a specific example?''.
Finally, participants discussed what capabilities an ideal tool should provide to better support molecular optimization workflows.
Participants were compensated with around US\$6 for their time.


\subsection{Findings}

All participants agreed that optimizing small molecules requires substantial effort.
They described it as an iterative design workflow consisting of several recurring stages.
In practice, they repeatedly modify candidate molecules through structural changes, such as functional-group substitutions or scaffold adjustments, using molecule design tools.
They then assess the resulting candidates with computational tools to estimate properties relevant to their design goals, such as activity, toxicity, and synthesizability.
In parallel, they consult additional evidence sources, including database search, literature comparison, and patent checks, to inform both the assessment of current candidates and the direction of subsequent modifications.
Throughout this process, experts repeatedly move between structure editing, property evaluation, and evidence gathering while coordinating multiple tools.
Within this workflow, participants increasingly leverage GenAI systems to support molecular optimization, primarily to suggest structural modifications, explore alternative design directions, and accelerate early-stage ideation.
However, they identified several challenges when collaborating with these GenAI systems in molecular optimization.

{\textbf{C1: Chemists struggle to express multifaceted modification intent to guide AI-assisted candidate generation.}}
Participants (E1, E3, and E4) often had clear intent regarding how to modify a candidate molecule, based on their experience and domain knowledge.
E3 stated, ``I found a certain ring is toxic, and I explicitly told the model that this ring should be replaced with other functional groups.''
Across interviews, we found that such modification intentions were often multifaceted, spanning structural constraints (E1, E3), desired property changes (E1, E4), and task-level optimization goals  (E4, E6).
For example, participants may want to modify a particular functional group while preserving the core scaffold, improve properties such as activity, toxicity, or synthesizability, or steer optimization toward broader therapeutic objectives. 
However, existing GenAI systems primarily rely on natural-language input and provide limited support for expressing such multifaceted intent precisely.
As a result, participants often had to resort to cumbersome workarounds, such as annotating molecular images with external tools or manually referring to target regions through textual molecular representations, both of which required additional translation effort and remained prone to ambiguity.
The difficulty was further amplified when participants needed to specify multifaceted intent, as communicating multiple design requirements made annotation and prompt construction even more complex (E1, E4).

{\textbf{C2: Chemists face challenges in evaluating and tracing AI-generated recommendations.}}
Participants noted that, although GenAI systems can rapidly generate candidate molecules, substantial additional effort is still required to determine whether these recommendations are scientifically meaningful and worth following.
They described current GenAI systems as largely opaque and black-box in nature, offering little explanation or supporting evidence for their outputs.
E5 noted, ``We do not trust the models because they are black boxes.'' 
This lack of transparency made it difficult for participants to assess the novelty of generated candidates (E1), judge the reliability of the recommendations (E3, E5), and justify adopting them in subsequent design decisions (E1, E5).
E5 further noted that recommended molecules require ``extremely strong proof,'' because ``collaborators would otherwise be reluctant to synthesize or follow up on AI-generated suggestions.'' 
As a result, they often had to conduct substantial follow-up verification, such as consulting external tools, searching the literature, and relying on their own expertise (E1, E3, E4). 
In addition, participants further noted that current GenAI interactions are typically presented as linear chat histories or isolated outputs, which makes it difficult to trace how a molecule evolves across iterations, revisit earlier alternatives, or compare different modification directions. 
This hindered their ability to inspect AI-generated changes and maintain decision context.



{\textbf{C3: Chemists face frequent and tedious tool switching during molecular optimization.}}
Participants {(E1, E3, E4, E5, E6)} noted that current GenAI tools are not well designed to support the molecular optimization workflow as a whole.
In practice, they mainly used GenAI tools to suggest candidate modifications or propose new molecular structures, while relying on separate platforms for molecule visualization and refinement {(e.g., Prompt LM and MOE)}, property prediction {(e.g., artic, Swiss, and ADMET 3.0)}, and database or literature search {(e.g., PubChem and USPTO)}, because these downstream tasks required more specialized functions and more reliable computational or evidence support.
As a result, participants had to frequently switch between tools to inspect, validate, and refine AI-generated candidates. 
This fragmented workflow required required participants to spend additional effort coordinating different tools and information sources, which disrupted their reasoning during iterative optimization.
For example, E1 described this overhead, ``To compute different properties, we need different tools... switching back and forth is time-consuming, such as using one tool for logP and another for toxicity, and then returning to GPT for the next step.'' 
E5 similarly noted that ``different properties require different tools,'' highlighting the need to coordinate across multiple platforms.



\subsection{Design Requirements}
\label{sec:design_re}
We derive five design requirements for interactive LLM-assisted molecular optimization systems according to the findings.

\textbf{DR1: Offering easy expression of users' multifaceted modification intent  for guiding generation.}
The tool should allow users to externalize their modification ideas or domain knowledge in a clear and low-effort manner \textbf{(C1)}, such as preserving or editing specific substructures, specifying editable regions, referencing target structures from examples, and indicating desired property targets or broader optimization goals. 
We organize such intent into four complementary perspectives: high-level optimization goals, property-oriented preferences, structure-level editing intent for the current molecule, and example-based preference derived from other molecules.
With the tool, users can easily guide the candidate generation to align with their intent and knowledge.

\textbf{DR2: Generating intent-aware and chemically feasible modification recommendations.}
The tool should generate candidate molecules that respect user-specified modification intent \textbf{(C1)} while also considering fundamental chemical knowledge, such as structural plausibility, so that the recommendations are both feasible and aligned with user intent.

\textbf{DR3: Providing evidence and provenance for each recommendation.}
The tool should provide explanations and verifiable evidence for each recommendation \textbf{(C2)}, such as links to relevant literature, database records, and prior-art checks.
With such support, users can better assess the credibility of AI-generated suggestions and make informed molecular design decisions.

\textbf{DR4: Supporting traceable and comparable AI-assisted iterations.}
The system should preserve the history of molecular modifications to make AI-assisted iterations easy to trace, inspect, and compare \textbf{(C2)}. 
With this design, users can review each modification step, compare alternative design directions, and revisit earlier candidates when needed.

\textbf{DR5: Supporting AI-assisted molecule optimization in an integrated environment.}
The system should bring together GenAI-based candidate generation, chemical property computation platforms and commonly used functions for molecular visualization and refinement, property prediction, and evidence lookup \textbf{(C2, C3)}. 
With such integration, users can reduce tool switching and more efficiently inspect, validate, and refine AI-generated candidates.




\section{\tool}
In this section, we first present an overview of \tool (\autoref{sec:sys_overview}). 
We then describe the interface design of each panel (\autoref{sec:interface_design}) and the computational pipeline that processes user input and generates candidate molecules (\autoref{sec:computational_pipeline}).

\begin{figure*}[t]
    \centering
    \includegraphics[width=0.95\linewidth]{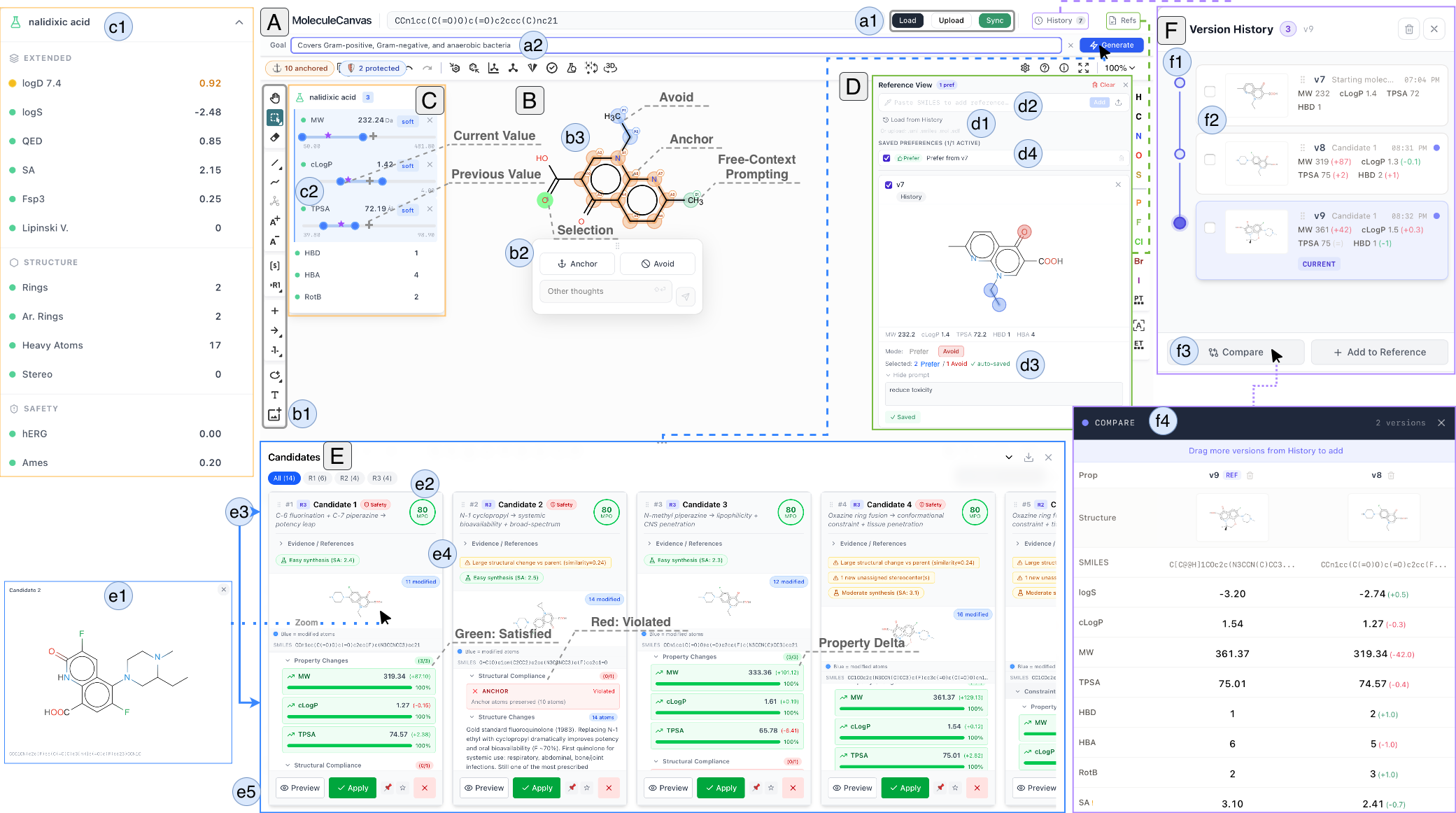}
    \caption{The interface of \tool supports iterative molecular optimization through coordinated panels. (A) specifies high-level optimization goals and initializes the molecule. (B) enables structure-level editing with anchor, avoid, and in-context prompting to express modification intent. (C) supports property-oriented intent by exposing current values and target ranges. (D) externalizes and reuses reference-based structural preferences from prior candidates or examples. (E) presents generated candidates with scores, intent satisfaction, and supporting evidence for inspection and selection. (F) maintains a branching history of design iterations, enabling tracing, comparison, and revisiting of alternative optimization paths.}
    \Description{This figure shows the MolecularCanvas interface, organized around a large central molecule-editing canvas and divided into six coordinated panels labeled A through F. Panel A forms a horizontal task and navigation bar across the top. Panel B occupies most of the central area beneath it. Panel C appears as a narrow property panel along the left side of Panel B, while Panel D is positioned along its right side. Panel E spans the lower portion of the interface and contains a horizontal row of candidate cards. Panel F occupies the far-right column, with the version-history view at the top and a comparison table extending below it. Circular callouts labeled a1 through f4 mark specific controls and interface elements within these panels. Panel A serves as the top-level task and navigation bar for MolecularCanvas. It contains the SMILES input field and controls for loading a molecule, uploading a molecular file, and synchronizing the current structure (a1). It also contains a natural-language field for specifying the high-level optimization goal (a2). On the right, the Refs, Generate, and History buttons open the reference view in Panel D, trigger the optimization process and display generated candidates in Panel E, and open the version-history view in Panel F, respectively. Panel B displays the current molecular structure on a large editing canvas. A vertical toolbar along the left edge provides standard molecular editing operations (b1). When users select an atom or substructure, a contextual menu appears beside the molecule, allowing them to mark the selected region as an Anchor or Avoid constraint or enter a localized free-text instruction (b2). The resulting annotations are rendered directly on the molecular structure: orange outlines mark anchored regions, blue highlights mark regions to avoid modifying, dark green highlights indicate the location associated with a free-context instruction, light green highlights the current selection (b3). Panel C presents molecular properties and their target settings of the current molecular in Panel B. The main property list groups descriptors into categories such as extended physicochemical properties, structural characteristics, and safety indicators, with each row showing the property name, current value, and a colored status marker (c1). Expanding selected properties opens a detailed view containing descriptors such as molecular weight, cLogP, and TPSA (c2). For each property, a horizontal slider shows the allowable range and uses distinct markers to represent the current value, previous value, and optimization target. Users can adjust the target directly on the slider. Panel D presents reference-based preferences in a vertical reference-management view. At the top, users can load a molecule from the design history (d1) or add a new reference by entering a SMILES string or uploading a molecular file (d2). Saved references appear below as expandable entries, each with a checkbox for activating or deactivating that reference during generation (d4). The expanded entry displays the reference molecule as a two-dimensional structure with colored highlights marking preferred and avoided regions. A summary beneath the structure reports the selected regions, and a text field allows users to attach an additional free-form preference, such as “reduce toxicity” (d3). After the user specifies the optimization intent and clicks Generate button at Panel A, Panel E displays the generated molecules as horizontally arranged, ranked candidate cards. Within each card, information is organized vertically from a brief candidate summary at the top to detailed evaluation results and action controls at the bottom. Each card contains a 2D structure preview (e1), an overall optimization score (e2), summaries of satisfied and violated structural and property requirements (e3), and supporting evidence (e4). Satisfied requirements appear in green and violated requirements in red. There are four buttons at the bottom of each card (e5) allow users to preview, retain, reject, or apply a candidate.Panel F records the optimization process as a branching version history. A vertical timeline on the left (f1) connects successive versions, while each version card displays a molecule thumbnail, version number, source, timestamp, and selected properties, such as molecular weight, cLogP, TPSA, and hydrogen-bond donors. A checkbox beside each card (f2) allows users to select multiple versions. The Compare button at the bottom (f3) opens a detailed comparison view, while Add to Reference button saves a selected version as a reference molecule. The comparison view (f4) presents the selected versions side by side in a table, showing their molecular structures, SMILES strings, and property values. Differences relative to the reference version are displayed in parentheses, with green and red indicating favorable and unfavorable changes, respectively.
}
    \label{fig:system_overview}
\end{figure*}


\begin{figure}[t]
  \centering
  \begin{minipage}{\columnwidth}
\begin{lstlisting}[style=compactjson]
{
  "SMILES": "CCn1cc(C(=O)O)c(=O)c2ccc(C)cnc21",
  "Goal": "Cover Gram-positive and anaerobic bacteria",
  "intent": {
    "property_constraints": [
      {
        "property_key": "MolWt",
        "current_value": 395.67,
        "target_value": [320, 380]
      },
      {
        "property_key": "LogP",
        "current_value": -1.86,
        "target_value": [1.0, 3.0]
      },
      {
        "property_key": "TPSA",
        "current_value": 176.95,
        "target_value": [60, 150]
      }
    ],
    "structure_constraints": [
      {
        "molecule": "current",
        "anchor_ids": [12, 13, 14],
        "avoid_ids": [27, 28],
        "free_form": {
          "pos": [23, 45],
          "prompt": "reduce toxicity"
        }
      }
    ]
  },
  "edit_plan": {
    "edit_groups": [
      {"pos": [23], "operation": "replace", "target": "F"},
      {"pos": [45], "operation": "add", "target": "polar_group"}
    ]
  }
}
\end{lstlisting}
  \end{minipage}
  \caption{Representative structured backend input and generated edit
  plan. The input contains the initial molecule and goal, property
  constraints, and atom-level structural constraints. The generated
  plan expresses the proposed modifications as executable operations.}
  \Description{This figure shows a vertically arranged, syntax-highlighted JSON example containing structured backend input and a generated edit plan. The JSON is organized from the initial molecule and goal at the top, through a structured intent object in the middle, to an edit plan at the bottom. Blue denotes top-level and major nested keys, purple denotes lower-level field keys, orange denotes text values, and green denotes numeric values, atom identifiers, and ranges. The initial molecule is represented as a SMILES string, with the goal “Cover Gram-positive and anaerobic bacteria.” The intent object records current and target values for three molecular properties: MolWt from 395.67 toward 320–380, LogP from −1.86 toward 1.0–3.0, and TPSA from 176.95 toward 60–150. It also anchors atom IDs 12, 13, and 14; marks atom IDs 27 and 28 as positions to avoid; and associates the instruction “reduce toxicity” with atom positions 23 and 45. The edit plan contains two executable operations: replacing the group at position 23 with fluorine and adding a polar group at position 45.}
  \label{fig:structure_input}
\end{figure}

\subsection{System Overview} 
\label{sec:sys_overview}

Based on the design requirements identified in~\autoref{sec:design_re}, we developed \tool, an interactive system for AI-assisted molecule optimization. 
As shown in Figure~\ref{fig:system_overview},
\tool includes six coordinated panels that support different aspects of iterative molecule optimization within a unified environment (\textbf{DR5)}.

Rather than relying on a single free-form prompt, \tool allows users to iteratively construct optimization context by combining high-level goals, structure-level annotations, property constraints, and reference-based preferences, which are then used to guide candidate generation (\textbf{DR1}).
The workflow typically begins in the \textit{Goal Specification Panel} (A), where users import an initial molecule and specify high-level optimization objectives, such as improving bioavailability or reducing toxicity (\textbf{DR1}).
The molecule is then synchronized with the central \textit{Molecule Canvas} (B), which serves as the primary workspace for inspecting and editing the molecular structures.
Our system supports the explicit expression of users’ multifaceted modification intent, spanning high-level goals, structural constraints, property targets, and contextual preferences, to guide candidate generation.
First, users can directly annotate selected substructures on the canvas to express structural intent, such as preserving, avoiding, or freely prompting how specific regions should be modified.
Furthermore, to complement structure-level editing, the \textit{Property Panel} (C) displays the predicted properties of the current molecule and allows users to specify desired property ranges or optimization targets, helping them express property-oriented intent. 
In addition, the \textit{Reference Panel} (D) allows users to reference other molecule examples and annotate useful patterns or preferred structures, further externalizing their design preferences and domain knowledge.

Based on these multifaceted modification intent, \tool generates molecule candidates in the \textit{Candidate Panel} (E) (\textbf{DR2}). 
Each candidate includes an overall score, summaries of how well it satisfies user intent, and supporting evidence, allowing users to inspect, compare, and validate AI-generated recommendations (\textbf{DR2}, \textbf{DR3}, \textbf{DR5}).
Finally, the \textit{History Panel} (F) records previously explored molecules and design steps, allowing users to revisit earlier states and compare candidates over time (\textbf{DR4}). 


\subsection{Interface Design} \label{sec:interface_design}

This section introduces the six panels of \tool in detail, including support for multifaceted intent expression, candidate exploration and evaluation, and iteration tracing during AI-assisted molecular optimization.

\subsubsection{Multifaceted Intent Expression}
\label{Multifaceted_intent}
Guided by \textbf{DR1}, \tool supports four complementary forms of intent expression through four dedicated panels: 
the \textit{Goal Specification Panel} for high-level optimization goals, the \textit{Canvas Panel} for structure-level editing intent on the current molecule, the \textit{Property Panel} for property-oriented preferences, and the \textit{Reference Panel} for example-based preferences derived from other molecules.


\textit{Goal Specification Panel}
allows users to initialize the optimization process by providing a lead molecule and defining the overall design objective.
Users can import a molecule by pasting a SMILES string or uploading an external structure file (a1). The molecule is then synchronized with the Canvas Panel, where users can further edit the structure using a built-in molecular editor.
Once the initial molecule is loaded, users specify the optimization goal through natural-language descriptions, such as reducing medication-related risks in specific usage scenarios (e.g., mitigating alcohol-related adverse reactions in cephalosporin-like compounds) (a2).



\textit{Canvas Panel} serves as the central workspace where users visualize and manipulate the 
molecular structure. 
The canvas integrates the open-source molecular editor \textit{Ketcher} \cite{ketcher}, enabling users to inspect the molecule and perform structure-level editing.
Users can interact with atoms, fragments, or substructures through standard molecular editing operations (b1).
Beyond manual 
editing, the canvas also allows users to select regions of the molecule and attach contextual annotations that describe intended modifications for AI-assisted optimization.
Specifically, users can annotate selected substructures as \textit{Anchor} regions to preserve them during optimization, label regions as \textit{Avoid} to discourage modifications in those areas, or attach free-form in-context prompts that describe desired modification strategies, such as functional-group substitutions or preferred chemical transformations (b2).
We provide dedicated \textit{Anchor} and \textit{Avoid} controls because these two operations were frequently raised in the formative study, whereas free-form prompts support other, less standardized modification intents.
Selected regions are visually highlighted with color-coded overlays that appear directly on the molecular structure, allowing users to 
distinguish regions associated with different annotations (b3).
These structure-grounded annotations turn prompting from a purely textual instruction into a {structure-aware in-context interaction}, allowing users to specify desired modifications directly in relation to molecular regions rather than describing them in isolation.


\textit{Property Panel} serves two purposes: it allows users to inspect the properties of the current molecule and to specify property-level optimization intent.
It includes common chemical properties, such as cLogP, MW, and TPSA, which are equipped with literature-informed ranges and organized into five categories based on medicinal chemistry criteria: physicochemical, solubility, ADME, safety, and developability \cite{lipinski2012experimental, veber2002molecular, bickerton2012quantifying} (c1).
Users can select which molecular properties should be considered during optimization by clicking to expand individual properties and defining their desired target ranges using interactive sliders, thereby shaping the search space for candidate generation (c2).
The slider shows both the property value of the current molecule and that of the previous iteration, helping users adjust optimization targets while maintaining awareness of how the molecule evolves across design rounds.


\textit{Reference Panel} allows users to externalize and reuse structure-level preferences derived from other molecules, either selected from previously explored candidates (d1) or newly uploaded examples (d2). 
While the \textit{Molecule Canvas Panel} captures users' intended modifications for the current molecule, the \textit{Reference Panel} supports reference-based preference expression, enabling users to borrow useful substructures or design ideas from prior examples.
In this panel, users can specify similar types of structural preferences as in the \textit{Canvas Panel}, including \textit{Prefer}, \textit{Avoid}, and free-form prompts (d3), which are then used to guide later optimization rounds.
Users can activate or deactivate each created reference-based preference by checking or unchecking the corresponding checkbox (d4).



\subsubsection{Candidate exploration and evaluation.}

\textit{Candidate Panel}
presents AI-generated molecules as a ranked list of candidate cards to support exploration and evaluation.
Each card contains a 2D structure preview (e1), an overall optimization score (e2), a summary of how well the candidate satisfies user intent (e3), and supporting evidence (e4).
To help users quickly evaluate candidates, the intent-satisfaction summary is organized from two perspectives: structural and property-oriented requirements. 
Satisfied requirements are highlighted in green, whereas unmet ones are highlighted in red (e3). 
The evidence section 
explains why a candidate is recommended by summarizing its key changes relative to the current molecule and presenting relevant external support from public pharmaceutical databases (ChEMBL and PubChem), including known-compound matches, safety signals such as drug withdrawals and black-box warnings, clinical trial phases of structurally similar drugs, and related bioassay and patent records, for deeper evaluation (e4).
After reviewing, users can then preview, retain, reject, or apply a candidate as the starting point for the next optimization round (e5).


\subsubsection{Iteration tracing.}

\textit{History Panel}
maintains a record of previously explored molecules and design steps throughout the optimization process.
The panel visualizes optimization trajectories as a branching sequence of parent–child states, allowing users to inspect how candidate molecules evolved across iterations (f1).
Users can revisit earlier molecules, review candidates generated in prior rounds, and restore previous molecules as starting points for further exploration.
Users can select multiple versions by clicking the corresponding checkboxes (f2) and click \textit{Compare} {(f3)} to open a side-by-side comparison view, where property differences and structural changes across the selected versions are displayed together (f4). 


\subsection{Computational Pipeline} 
\label{sec:computational_pipeline}

We developed a computational pipeline that generated a ranked list of candidate molecules aligned with users' multifaceted modification intent.  
The pipeline consists of three stages: (1) formalizing users’ modification intent from their interface inputs, (2) generating candidate molecules conditioned on this intent, and (3) evaluating and ranking the generated candidates.


\subsubsection{Modification Intent Formulation}
We propose a formal specification of users’ multifaceted modification intent, as illustrated in {\autoref{fig:structure_input}}. 
This specification contains two components. 
The first is the molecule to be optimized, represented in SMILES format, which is typically provided through the \textit{Canvas Panel} or the \textit{Goal Specification Panel}. 
The second component is the modification intent, which includes three aspects: (1) the optimization goal specified as text in the \textit{Goal Specification Panel}; (2) property constraints from the \textit{Property Panel}, represented as a list of target properties with their \textit{name}, \textit{current value}, and \textit{target value}; and (3) structural constraints from the \textit{Canvas Panel} and the \textit{Reference Panel}, where each entry contains a molecule in SMILES format, and a list of annotated substructure with explicit atom IDs that define their positions. Each annotation is associated with an intent label such as \textit{anchor}, \textit{avoid}, \textit{prefer}, and free-form modification prompts. \revise{User selections are captured directly from the molecular editor as atom and bond identifiers, without requiring vision–language interpretation of the rendered molecular structure.}


\subsubsection{Candidate Generation}
Given the structured modification intent, we adopt a two-step generation strategy. 
Rather than generating complete molecules directly, we first use a GenAI model \revise{(GPT-4o in our experiments)} 
to produce an explicit edit plan, and then apply the plan to the target molecule to generate candidate structures.
We adopt this design because high-level optimization goals and property constraints are realized through structural modifications, which may conflict with user-specified structural constraints. 
Direct molecule generation makes such conflicts difficult to detect and may produce candidates that violate users’ 
structural intent. 
By introducing an intermediate edit plan, we make proposed structural changes explicit before execution, so they can be checked against structural constraints and discarded if conflicts arise.

For edit-plan generation, we send the structured intent representation to the LLM and instruct it to produce structural edit operations. 
As shown in Figure ~\ref{fig:structure_input}, each operation specifies where to modify the molecule, what transformation to apply (e.g., add, remove, or replace), the corresponding material needed for the transformation, such as \textit{replacing} a \textit{chlorine atom} at a {target position} with \textit{fluorine}. 
This operation format is designed to be directly executable by the reaction engine RDKit \cite{rdkit} for generating concrete candidate molecules. 

The system then executes the edit plan by translating each operation into a molecular transformation using RDKit.
For \textit{replace} and \textit{add} operations, the system locates the target fragment on the molecule, substitutes or attaches the specified replacement, and preserves the surrounding structure.
For \textit{remove} operations, the matched fragment is deleted, and the largest remaining connected structure remains.

We then validate each generated candidate in two steps.
First, we check chemical validity, ensuring that the molecule is structurally sound with correct valence, aromaticity, and ring-bond consistency. 
Second, we remove exact duplicates by canonical SMILES comparison and near-duplicates by Morgan fingerprint similarity (Tanimoto coefficient \(> 0.85\)) to maintain structural diversity.

\revise{This two-step generation strategy separates LLM-based edit-plan generation from RDKit-based execution and validation.
Although the choice of LLM may affect edit-plan quality and executability, the validation procedures are model-agnostic and ensure that only valid candidates satisfying the constraints are presented to users.
}

\subsubsection{Candidate Evaluation}
After candidate generation, we score each candidate along three dimensions: property constraint satisfaction, structural constraint satisfaction, and chemical plausibility. 
We do not directly score the high-level optimization goal or prompt-based structure constraints, as they mainly guide generation and lack reliable post hoc evaluation metrics.
Specifically, for property satisfaction, we compute the molecular properties for each candidate and assign scores based on whether they satisfy the specified target ranges. 
For structural satisfaction, we check whether user-specified structural constraints, such as anchored and avoided substructures, are respected, and assign scores accordingly.
For chemical plausibility, we assess each candidate using established drug-likeness heuristics, including Lipinski’s Rule of Five and Veber’s rule, with violations incurring score penalties. Unlike the validity check during generation, which discards molecules that are chemically impossible, this step evaluates whether a valid molecule is likely to be viable as a drug candidate.

We then rank the candidates based on the aggregated scores and select the top-\(k\) molecules for user exploration. 
For each selected candidate, we further retrieve supporting evidence from databases such as ChEMBL, including structurally similar known compounds (Tanimoto similarity \( \geq 0.8 \)), exact matches to existing drugs, associated safety signals (e.g., withdrawn status and black-box warnings), clinical-phase information, and bioactivity data.

\section{Evaluation}
\begin{figure*}[t]
    \centering
    \includegraphics[width=0.94\linewidth]{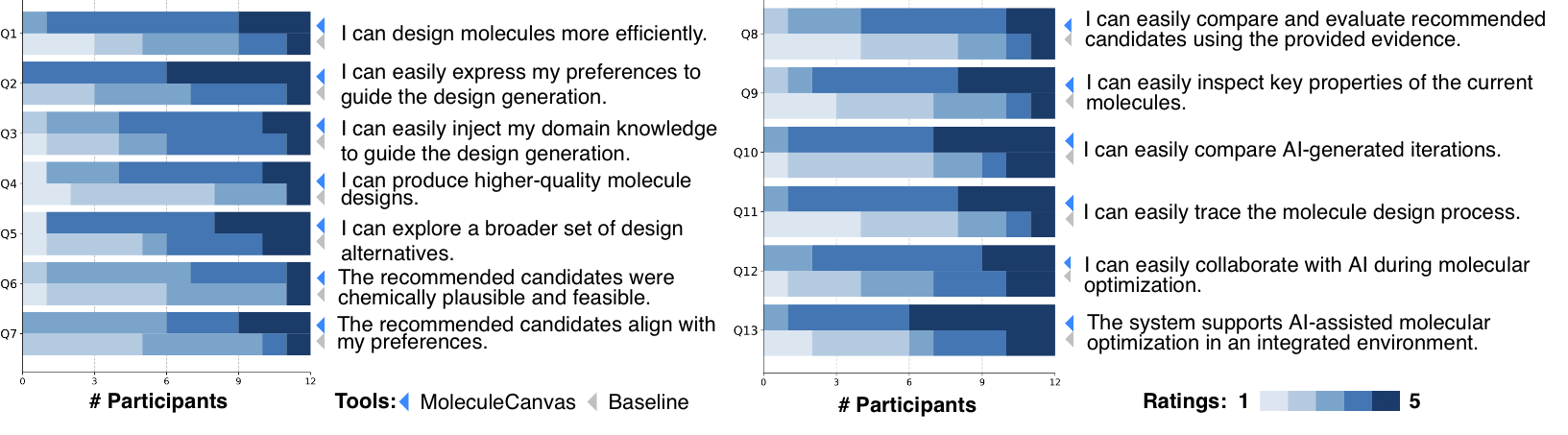}
    \caption{Participants’ ratings for \tool and the baseline across Q1--Q13. Each item shows paired distributions (top: \tool, bottom: baseline). \tool is rated significantly higher than the baseline on 13 questions (p < 0.05).}
    \Description{A two-column horizontal stacked-bar chart shows Q1--Q7 on the left and Q8--Q13 on the right. Each question has two bars each totaling 12 participants: MolecularCanvas above, marked by a blue triangle, and the baseline below, marked by a gray triangle. Question statements appear beside each bar pair, and the horizontal axes show participant counts from 0 to 12. Within each bar, ratings run from 1 on the left to 5 on the right and are encoded from pale to dark blue. Across all thirteen questions, MolecularCanvas bars contain more responses in the dark-blue ratings 4 and 5, whereas baseline bars contain more pale and medium-blue ratings 1 through 3. Particularly large visual differences appear for evidence-supported candidate evaluation, inspection of molecular properties, comparison of generated iterations, and tracing the design process.}
    \label{evaluation}
\end{figure*}

We conducted a user study with 12 participants to evaluate the effectiveness and usability of \tool.
We introduce the study setup in \autoref{study_setup} and the results~\autoref{results}.

\subsection{Study Setup}
\label{study_setup}

\subsubsection{Participants}\label{participants}
In our user study, we recruited 12 participants (8 males and 4 females, Aged $28.6\pm 3.96$) from academic referral, social media advertisement and targeted outreach within medicinal chemistry communities.
They included 7 industry researchers/scientists, 3 PhD students, and 2 Master’s students, with primary expertise spanning medicinal chemistry (N=6) and computational chemistry (N=6).
They have experience in {chemistry-related molecular design and drug discovery}, with $4.50 \pm 3.50$ years. Most participants (9/12) had prior experience with AI-based molecular design tools.

\subsubsection{Baseline.}
As no directly comparable system was available that integrates candidate generation, property evaluation, and molecule visualization and editing in a single platform,
participants were allowed to use any tools they were familiar with in the baseline conditions.
These tools typically included conversational AI tools for molecular ideation and candidate suggestion (e.g., Google Gemini, Claude, and OpenAI ChatGPT), property evaluation tools for assessing molecular properties {(e.g., RDKit, ADMETlab, SwissADME)}, and molecule visualization and editing tools for structure inspection and modification {(e.g., MolView, ChemDraw, Avogadro)}.

\subsubsection{Tasks and Materials}\label{task}
We prepared one warm-up task and two formal tasks on antibiotic molecular optimization.
In all tasks, participants were asked to iteratively modify a given starting molecule under multiple design goals and constraints, based on their own reasoning and domain judgment.
All tasks were open-ended, with no single correct answer, to reflect the exploratory and iterative nature of real-world molecular optimization. \revise{These tasks were intended to evaluate early-stage design exploration rather than the full small-molecule optimization cycle involving experimental validation.
} The warm-up task used Penicillin G as a practice example to help participants become familiar with the system.
Participants were asked to improve its oral availability, broaden its antibacterial spectrum, and increase its \(\beta\)-lactamase resistance while preserving the \(\beta\)-lactam core.
The two formal tasks were designed around realistic antibiotic optimization scenarios.
Task A used Nalidixic acid as the starting molecule for optimization toward a modern quinolone with broader antibacterial coverage, appropriate drug-like properties, safety, and a high resistance barrier.
Task B used Cefazolin as the starting molecule for optimization toward a broad-spectrum cephalosporin effective against both Gram-positive and Gram-negative bacteria, including MRSA, while improving \(\beta\)-lactamase stability and satisfying given property constraints.
More details of the tasks and materials are provided in the supplementary material.


\subsubsection{Procedure}\label{procedure}
Each study session lasted approximately 1.5 hours and was conducted as a one-on-one online session.
At the beginning of each session, we introduced the study procedure and obtained participants' consent to record the session for research purposes.
Participants then completed the study tasks using two systems across two tasks in a counterbalanced order to reduce order effects.
Before using \tool, participants received a brief tutorial introducing its main components and functionalities, and then completed a warm-up task to familiarize themselves with the system.
The tutorial phase ended once participants indicated that they felt sufficiently familiar with \tool, which typically took about 10 minutes.
For each task, participants could stop once they were satisfied with their final molecule or when the 20-minute time limit was reached.
After using each system, participants completed post-condition questionnaires, including the System Usability Scale (SUS)~\cite{brooke1996sus} and additional questionnaires assessing perceived system effectiveness {(\autoref{evaluation})}, while following a think-aloud protocol.
All questionnaire items were rated on a 5-point Likert scale, ranging from 1 (\textit{strongly disagree}) to 5 (\textit{strongly agree}).
At the end of the session, we conducted a semi-structured interview to discuss the strengths and weaknesses of both conditions, how \tool influenced participants' workflows, and potential improvements to \tool.
Each participant received approximately US \$8 as compensation for their time and effort.

\subsection{Results}\label{results}
In this section, we report the quantitative and qualitative results of the user study.

\subsubsection{Quantitative results}
We first report the questionnaire results on perceived effectiveness and usability. We then complement these findings with an expert evaluation on overall quality of the final molecules produced by participants.
\revise{Finally, we assess the reliability of the generation pipeline using logs collected during the user study.}


\textbf{Effectiveness.}
As shown in \autoref{evaluation}, \tool received significantly higher ratings than the baseline across all measured aspects (all \(p<0.05\)).
Overall, participants rated \tool significantly higher in supporting efficient molecule design (Q1, \(4.17 \pm 0.58\) vs. \(2.67 \pm 1.30\), \(p<0.01\)). 
We further organize the remaining results into four perspectives: modification intent expression, molecule generation quality, evaluation and inspection, and workflow support.

    For \textit{modification intent expression}, \tool significantly outperformed the baseline in helping users express preferences (Q2, \(4.50 \pm 0.52\) vs. \(3.25 \pm 0.97\), \(p<0.01\)) and inject domain knowledge (Q3, \(3.75 \pm 0.87\) vs. \(3.17 \pm 1.19\), \(p<0.05\)) to guide molecule design.

For \textit{molecule generation quality}, participants rated \tool significantly higher in generating high-quality (Q4, \(3.67 \pm 1.07\) vs. \(2.33 \pm 1.07\), \(p<0.01\)), diverse (Q5, \(4.08 \pm 1.08\) vs. \(3.17 \pm 1.34\), \(p<0.05\)), chemically plausible and feasible (Q6, \(3.42 \pm 0.79\) vs. \(2.58 \pm 1.00\), \(p<0.01\)), and preference-aligned candidates (Q7, \(3.75 \pm 0.87\) vs. \(2.83 \pm 0.94\), \(p<0.05\)).

For \textit{evaluation and inspection}, \tool received significantly higher ratings for supporting users in evaluating recommended candidates with evidence (Q8, \(3.75 \pm 0.87\) vs. \(2.25 \pm 1.29\), \(p<0.05\)), inspecting key properties of current molecules (Q9, \(4.08 \pm 0.90\) vs. \(2.42 \pm 1.24\), \(p<0.01\)), comparing AI-generated iterations (Q10, \(4.33 \pm 0.65\) vs. \(2.75 \pm 1.29\), \(p<0.01\)), and tracing the overall molecule design process (Q11, \(4.25 \pm 0.62\) vs. \(2.25 \pm 1.29\), \(p<0.01\)).

For \textit{workflow support}, participants rated \tool significantly higher in enabling collaboration with AI (Q12, \(4.08 \pm 0.67\) vs. \(3.17 \pm 1.27\), \(p<0.05\)) and integrating AI into their molecule optimization workflows (Q13, \(4.42 \pm 0.67\) vs. \(2.92 \pm 1.44\), \(p<0.05\)).

\textbf{Usability.}
\tool received a higher SUS score than the baseline (\tool: \(72.50 \pm 10.82\), Baseline: \(60.83 \pm 14.16\); Wilcoxon \(p < 0.05\)). 
A SUS score of 72.5 suggests good usability~\cite{sauro2016quantifying}. 
More detailed ratings are provided in the supplementary material.


\textbf{Overall quality of the designed molecules.}
We invited domain experts in medicinal chemistry, with 3 and 7 years of experience, respectively, to rate the final molecules from each condition based on overall design quality. 
Experts were given the design goal of each task and asked to independently evaluate all molecules based on how well they align with these  objectives, considering aspects such as molecular validity, preservation of the core scaffold, similarity to known drug-like compounds, and overall structural plausibility.
To avoid bias, the source of each molecule was anonymized, and the molecules in each task were presented in random order during evaluation.
The results show that the molecules generated with \tool received higher ratings than those from the baseline (mean $0.73$ for \tool\ vs.\ $0.58$ for the baseline, on a normalized 0 to 1 scale), indicating improved overall design quality. This difference was statistically significant (Wilcoxon signed-rank test, $p < 0.001$). The two experts showed strong agreement, with an intraclass correlation coefficient (ICC) \cite{shrout1979intraclass} of $0.94$.

\revise{
\textbf{Generation reliability based on study logs.}
To further assess the reliability of the generation pipeline, we analyzed the backend logs collected during the user study.
Across all generation attempts, 95.0\% of LLM outputs produced executable edit plans, 57.0\% of generated candidates passed the validity and diversity filtering, and 91.0\% of the retained candidates satisfied the specified structural constraints.
Collectively, these rates suggest that roughly two generation attempts can yield one valid, non-duplicate, and constraint-satisfying candidate.
In practice, because \tool generates multiple candidates in parallel, the system can recommend enough valid candidates even when some generation attempts are filtered out.
}

\subsubsection{Qualitative Results}
We analyzed participants’ think-aloud protocols and post-task interviews to derive qualitative themes. All sessions were recorded and transcribed. One author coded the transcripts, focusing on how participants interacted with the system and how its features shaped their design process. The authors then met to review and consolidate the codes into higher-level themes. During the process, some themes were found to span multiple aspects of the interaction; in such cases, we assigned emphasis based on the aspects most clearly reflected in participants’ accounts. These themes are reported below with representative quotations as supporting evidence, complementing the quantitative results.

\textbf{\tool supports fine-grained and structure-level expression of design intent (9/12).} 
Participants consistently described using structure-level controls (e.g., anchor, avoid, and constraints) to directly specify which parts of a molecule should be preserved or modified. U1 noted that \textit{``anchor, avoid, and in-context prompt are very helpful when I'm trying to keep the quinolone core structure''}, while U12 explained that these operations allow them to \textit{``very directly define requirements and solve problems in a targeted way''}. 
These behaviors show that participants externalized domain knowledge through explicit structural operations rather than iterative prompt reformulation. Interaction logs further indicate that participants repeatedly used anchor operations and in-context prompts across iterations, suggesting that these mechanisms were actively used to guide localized modifications.

\textbf{\tool reduces cross-platform coordination by integrating key workflow steps (8/12).} 
Participants repeatedly contrasted \tool with baseline workflows that required switching across multiple tools. U1 described that in baseline settings they needed to \textit{``switch across platforms for visualization''}, while U3 noted that baseline workflows involved \textit{``jumping across platforms for visualization, property validation, and evidence search''}. In contrast, participants used \tool to perform structure editing, property inspection, and candidate evaluation within a single interface. U6 emphasized that this integration \textit{``greatly reduces the need to know how to manually modify structures''}, and U11 described the system as \textit{``clean and integrated''}. This consolidation enabled more continuous workflows without platform switching.

\textbf{\tool supports comparison-based decision making across candidate sets (10/12).} 
Rather than accepting a single generated molecule, participants described explicitly comparing multiple candidates to guide optimization decisions. U5 described that \textit{``I compare several generated molecules by looking at their property changes and structures, and then choose the one that best matches my goal.''} In practice, participants inspected candidate cards to compare property changes and structural differences before choosing which molecule to apply. U6 further explained that \textit{``in practice we select a group of promising molecules and compare them, rather than trusting a single generated result.''}.
Similarly, U9 highlighted that \textit{``version control makes the optimization process visible''}, allowing them to compare outcomes across iterations.

\textbf{\tool enables non-linear exploration through history-based backtracking (7/12).} 
Participants did not follow a strictly linear optimization trajectory, but instead revisited previous states and explored alternative modification strategies. U3 described that \textit{``after generating candidates, I go back to previous versions and try different modifications to see how the results change.''} U4 noted that \textit{``I use the history tracker to revisit previous versions and try different modification directions while interacting with the AI.''} In several cases, participants compared different iterations or returned to earlier molecules after observing undesirable changes, effectively exploring multiple branches of the design space. This supports counterfactual reasoning, where users contrast current results with alternative trajectories. 


\textbf{Suggestions.} In their suggestions for improvement, participants often framed \tool not as a standalone generator, but as the basis of a broader professional workflow. Some participants asked for richer contextual support, such as 3D structures, chemistry-specific evidence, or clearer accounts of how each candidate was generated. For example, U10 commented that \textit{``it would be better if the system could tell users how each candidate on the card was produced.''} U2 further suggested that integrating synthesis routes together with supporting evidence would make the system more practically useful, noting that \textit{``if synthesis routes and evidence are provided, it could be industrially useful.''} Overall, these suggestions point to opportunities to strengthen explanation, connect generated candidates to downstream decision-making, and better align the system with real-world medicinal chemistry workflows.



\section{Discussion}
This work explores how interactive system design can better support human-AI collaboration in molecule editing tasks. 
Based on our system design and user study findings, we reflect on broader design implications for AI-assisted scientific tools and discuss current limitations that point to promising directions for future work.

\subsection{Design Implications}
We summarize three design implications that were derived from the design process of \tool as follows.

\textbf{Direct structural interaction improves intent expression for prompting.}
\tool allows users to directly manipulate molecular structures to express modification intentions when prompting LLMs, instead of relying only on natural language. 
This addresses a central challenge identified in our formative studies: chemists face high cognitive load and ambiguity when trying to convey precise modification intentions through text alone, such as ``preserve this ring'' or ``modify this side chain''.
Our study shows that this approach reduces cognitive load and ambiguity, producing candidate molecules that better match user intentions.
This insight suggests a broader design principle: in domains with strong spatial, topological, or visual constraints, graphical and structure-based interactions should serve as the primary channel for intent communication.
In this way, the domain object’s structure can become a first-class entity. 
Users can directly annotate and constrain it while aligning with AI suggestions. This paradigm extends beyond molecular design to other structure-intensive domains such as protein engineering or materials science.

\textbf{Branching history and the reversibility of design reasoning.}
The History Panel in \tool supports users in tracing, comparing, and exploring alternative design branches. Beyond addressing the functional challenge of ``hard-to-track iterations'', this feature fundamentally reshapes how users perceive risk and approach exploration when collaborating with AI. Our study showed that when users are aware that they can revert to previous molecular states or compare alternative branches, they are more willing to pursue bold modifications, such as altering core scaffolds or introducing unconventional functional groups. This ``reversible design space'' lowers the psychological cost of trial-and-error, shifting optimization from a linear process to a graph-like exploration. Reversibility, therefore, should be seen not merely as an undo feature, but as a cognitive permission for experimentation. Future systems should treat branching comparison, state rollback, and branch merging as crucial capabilities rather than after-the-fact add-ons.
Moreover, the History Panel naturally aligns with the concept of a scientific notebook. Traditional workflows require chemists to manually document design decisions, rationales, and candidate outcomes, which is time-consuming and error-prone. Our iteration history automatically captures AI suggestions, user selections, and expressed intentions, forming a fully auditable and reproducible design reasoning chain. Future systems could further structure these histories to allow annotations, highlight key decisions, and support collaborative sharing, making ``reproducible design reasoning'' a core capability rather than a byproduct.

\textbf{Integrating external tools with LLM workflow.}
\tool integrates external computational tools (e.g., RDKit, SwissADME, ADMETlab) directly into the AI-assisted design workflow. Users can edit molecules and immediately see key properties (e.g., LogP, TPSA, QED, or synthetic feasibility) without switching platforms. This reduces workflow fragmentation and preserves cognitive continuity: users no longer need to remember structures, move data across tools, or rebuild context.
Our findings suggest that integration should focus on embedding capabilities into the core design task, rather than simply combining features in one interface. By seamlessly connecting generation and evaluation, systems let users focus on design reasoning instead of tool management. This can also extend to other scientific software, allowing flexible, on-demand integration of specialized tools while maintaining workflow continuity and supporting expert customization.

\subsection{Limitations and Future Works}
Despite the benefits demonstrated by \tool, our current design and implementation still have several limitations as follows.

\textbf{The accuracy of LLM generation in molecule-centric contexts.} 
Molecules themselves carry rich structural and functional information. Ensuring that the LLM truly understands users' prompts on molecules is essential for meaningful human–AI collaboration.
In our work, \tool employs a two-stage generation strategy to improve performance. 
First, our system will generate an edit plan, then execute molecular modifications based on that plan. 
However, the generation accuracy remains constrained by the underlying large language model’s (LLM) chemical knowledge coverage. Specifically, LLM reasoning is primarily based on textual molecular representations (e.g., SMILES), which pose inherent limitations when handling complex chemical information such as stereochemistry, chiral center retention or inversion, and conformational flexibility. 
In the future, we could incorporate multimodal generation models that combine molecular graphs, 3D conformations, and textual descriptions, enabling the system to handle richer chemical information. 

\textbf{The limitation of 2D-centered interaction in supporting users to express their intent.} 
\tool currently represents molecules and supports user interactions primarily in 2D, using SMILES strings and 2D structural renderings as the main formats. While this allows efficient visualization and manipulation of molecular topology, it cannot fully convey spatial information such as 3D conformations, binding modes with target proteins, or conformational flexibility. In some scenarios, these 3D features often critically influence design decisions: a modification that appears optimal in 2D may clash sterically or lose key interactions in 3D space.
To address this limitation, future systems could integrate 3D structural information into both interaction, visualization, and generation. For example, molecular docking and 3D visualization could allow users to inspect candidate molecules in the context of target protein pockets, annotate modifications based on spatial constraints, and guide the generative model to produce structures compatible with these 3D requirements. Such structure-aware interactions enable a more accurate representation of user intent.

\revise{\textbf{Support for requirement prioritization and property trade-offs.}
\tool allows users to express design intent through high-level goals, property targets, structural annotations, and reference-based preferences, but does not support explicit weighting or prioritization among these requirements. 
This limitation becomes particularly relevant when multiple property objectives conflict and users must make trade-offs among them.
Future work could introduce interactive controls for weighting and prioritizing requirements, together with analytical and visual support that reveals inter-property relationships, highlights potential conflicts, and helps users compare trade-offs among candidate molecules~\cite{Pajer2017WeightLifter, chen2026vizqstudio, sheng2026designpatterns, sheng2025trialcompass}.
}


\revise{\textbf{Extensibility to diverse laboratory workflows.}
\tool focuses on visual and LLM-based interaction for molecular optimization rather than serving as a general-purpose platform that supports the full range of cheminformatics functions.
Its current integrated workflow may therefore not fully accommodate the diversity and evolving nature of real laboratory practices. Future work could explore configurable integration mechanisms that allow users to incorporate external tools, models, and custom workflows~\cite{kim2023cells}.}

\revise{\textbf{Evaluation.}
We did not conduct a component-level ablation study, as a rigorous evaluation would require up to seven system variants and substantially longer sessions, imposing considerable burden on domain experts and making the study impractical and unmanageable. 
Future work should conduct targeted ablations to explore the contributions of individual components and longitudinal studies to evaluate sustained use and real-world impact.}

\section{Conclusion}
Molecular optimization is an iterative, constraint-driven process, yet existing GenAI tools provide limited support for expressing expert intent, inspecting model-generated modifications, and managing design iterations within a unified workflow. We presented \tool, an interactive system that integrates structure-level editing, constraint specification, GenAI-driven molecule generation, and evidence-based evaluation to support human–AI collaboration in molecular optimization.
Grounded in a formative study with medicinal chemistry experts, \tool addresses key challenges in current workflows by enabling fine-grained, structure-aware interactions, improving the transparency and traceability of generative recommendations, and reducing context switching across tools. Results from a user study show that \tool helps users better compare candidates, reason about trade-offs, and maintain coherent design trajectories when working with GenAI.
These findings highlight the importance of interactive, constraint-aware, and traceable GenAI systems for molecular optimization, and suggest broader design implications for human–AI co-design in scientific discovery.

\begin{acks}
We thank the reviewers and all study contributors.
This work is partially supported by the 
HK RGC GRF grant 16218724 
and by the
\grantsponsor{EU}{European Union}
{https://european-union.europa.eu/}
(ERC, SUSTECH, \grantnum{EU}{101224873}).
Views and opinions expressed are however those of the author(s) only and
do not necessarily reflect those of the European Union or the European
Research Council Executive Agency. 
Neither the European Union nor the granting authority can be held responsible for them.
\end{acks}

\bibliographystyle{ACM-Reference-Format}
\bibliography{main}

@ARTICLE{Pajer2017WeightLifter,
  author={Pajer, Stephan and Streit, Marc and Torsney-Weir, Thomas and Spechtenhauser, Florian and Möller, Torsten and Piringer, Harald},
  journal={IEEE Transactions on Visualization and Computer Graphics}, 
  title={WeightLifter: Visual Weight Space Exploration for Multi-Criteria Decision Making}, 
  year={2017},
  volume={23},
  number={1},
  pages={611-620},
  doi={10.1109/TVCG.2016.2598589}}

@ARTICLE{song2026vizdefender,
  author={Song, Sicheng and Zhang, Yanjie and Chen, Zixin and Qu, Huamin and Wang, Changbo and Li, Chenhui},
  journal={IEEE Transactions on Visualization and Computer Graphics}, 
  title={VizDefender: Unmasking Visualization Tampering Through Proactive Localization and Intent Inference}, 
  year={2026},
  volume={32},
  number={6},
  pages={4720-4730},
  doi={10.1109/TVCG.2026.3694448}}

@ARTICLE{song2025gvvst,
  author={Song, Sicheng and Zhang, Yipeng and Lin, Yanna and Qu, Huamin and Wang, Changbo and Li, Chenhui},
  journal={IEEE Transactions on Visualization and Computer Graphics},
  title={GVVST: Image-Driven Style Extraction From Graph Visualizations for Visual Style Transfer}, 
  year={2025},
  volume={31},
  number={9},
  pages={5975-5989},
  doi={10.1109/TVCG.2024.3485701}}

@article{chen2026vizqstudio,
  author={Chen, Zixin and Zeng, Yuhang and Song, Sicheng and Lin, Yanna and Xu, Xian and Qu, Huamin and Xia, Meng},
  journal={IEEE Transactions on Visualization and Computer Graphics}, 
  title={VizQStudio: Iterative Visualization Literacy MCQs Design With Simulated Students}, 
  year={2026},
  volume={32},
  number={8},
  pages={7468-7485},
  doi={10.1109/TVCG.2026.3695959}}

@article{lin2023inksight,
  title={Inksight: Leveraging sketch interaction for documenting chart findings in computational notebooks},
  author={Lin, Yanna and Li, Haotian and Yang, Leni and Wu, Aoyu and Qu, Huamin},
  journal={IEEE Transactions on Visualization and Computer Graphics},
  volume={30},
  number={1},
  pages={944--954},
  year={2023},
  publisher={IEEE}
}

@incollection{sauro2016quantifying,
title = {Chapter 9 - Six enduring controversies in measurement and statistics},
editor = {Jeff Sauro and James R. Lewis},
booktitle = {Quantifying the User Experience (Second Edition)},
publisher = {Morgan Kaufmann},
edition = {Second Edition},
address = {Boston},
pages = {249-276},
year = {2016},
isbn = {978-0-12-802308-2},
doi = {https://doi.org/10.1016/B978-0-12-802308-2.00009-6},
url = {https://www.sciencedirect.com/science/article/pii/B9780128023082000096},
author = {Jeff Sauro and James R. Lewis}
}

@article{hughes2011principles,
  author = {Hughes, JP and Rees, S and Kalindjian, SB and Philpott, KL},
  title = {Principles of Early Drug Discovery},
  journal = {British Journal of Pharmacology},
  volume = {162},
  number = {6},
  pages = {1239--1249},
  doi = {10.1111/j.1476-5381.2010.01127.x},
  year = {2011}
}

@article{waring2015analysis,
  title = {An Analysis of the Attrition of Drug Candidates from Four Major Pharmaceutical Companies},
  volume = {14},
  doi = {10.1038/nrd4609},
  number = {7},
  journal = {Nature Reviews Drug Discovery},
  author = {Waring, Michael J. and Arrowsmith, John and Leach, Andrew R. and Leeson, Paul D. and Mandrell, Sam and Owen, Robert M. and Pairaudeau, Garry and Pennie, William D. and Pickett, Stephen D. and Wang, Jibo and Wallace, Owen and Weir, Alex},
  year = {2015},
  pages = {475--486}
}

@article{papadatos2010mmp,
  author = {Papadatos, George and Alkarouri, Muhammad and Gillet, Valerie J. and Willett, Peter and Kadirkamanathan, Visakan and Luscombe, Christopher N. and Bravi, Gianpaolo and Richmond, Nicola J. and Pickett, Stephen D. and Hussain, Jameed and Pritchard, John M. and Cooper, Anthony W. J. and Macdonald, Simon J. F.},
  title = {Lead Optimization Using Matched Molecular Pairs: Inclusion of Contextual Information for Enhanced Prediction of hERG Inhibition, Solubility, and Lipophilicity},
  journal = {Journal of Chemical Information and Modeling},
  volume = {50},
  number = {10},
  pages = {1872--1886},
  year = {2010},
  doi = {10.1021/ci100258p}
}

@article{kenny2011mmp,
  author = {Griffen, Ed and Leach, Andrew G. and Robb, Graeme R. and Warner, Daniel J.},
  title = {Matched Molecular Pairs as a Medicinal Chemistry Tool},
  journal = {Journal of Medicinal Chemistry},
  volume = {54},
  number = {22},
  pages = {7739--7750},
  year = {2011},
  doi = {10.1021/jm200452d}
}

@article{sadybekov2023computational,
  title = {Computational Approaches Streamlining Drug Discovery},
  volume = {616},
  doi = {10.1038/s41586-023-05905-z},
  number = {7958},
  journal = {Nature},
  author = {Sadybekov, Anastasiia V. and Katritch, Vsevolod},
  year = {2023},
  pages = {673--685}
}

@article{lam2025navigating,
  title = {Navigating Structure-based Drug Discovery with Emerging Innovations in Physics- and Knowledge-based Approaches},
  volume = {2},
  doi = {10.1038/s44386-025-00031-4},
  number = {1},
  journal = {npj Drug Discovery},
  author = {Lam, Jordy Homing and Katritch, Vsevolod},
  year = {2025},
  pages = {29}
}

@misc{kingma2013vae,
      title={Auto-Encoding Variational Bayes}, 
      author={Diederik P Kingma and Max Welling},
      year={2013},
      eprint={1312.6114},
      archivePrefix={arXiv},
      primaryClass={stat.ML},
      url={https://arxiv.org/abs/1312.6114}, 
}

@inproceedings{jin2018junctiontree,
  title = 	 {Junction Tree Variational Autoencoder for Molecular Graph Generation},
  author =       {Jin, Wengong and Barzilay, Regina and Jaakkola, Tommi},
  booktitle = 	 {Proceedings of the 35th International Conference on Machine Learning},
  pages = 	 {2323--2332},
  year = 	 {2018},
  volume = 	 {80},
  publisher =    {PMLR},
  url = 	 {https://proceedings.mlr.press/v80/jin18a.html}
  
}

@article{chemformer2022,
  doi = {10.1088/2632-2153/ac3ffb},
  year = {2022},
  volume = {3},
  number = {1},
  pages = {015022},
  author = {Irwin, Ross and Dimitriadis, Spyridon and He, Jiazhen and Bjerrum, Esben Jannik},
  title = {Chemformer: A Pre-trained Transformer for Computational Chemistry},
  journal = {Machine Learning: Science and Technology}
}

@inproceedings{edwards2022translation,
  title = {Translation between Molecules and Natural Language},
  author = {Edwards, Carl  and
      Lai, Tuan  and
      Ros, Kevin  and
      Honke, Garrett  and
      Cho, Kyunghyun  and
      Ji, Heng},
  booktitle = {Proceedings of the 2022 Conference on Empirical Methods in Natural Language Processing},
  year = {2022},
  doi = {10.18653/v1/2022.emnlp-main.26},
  pages = {375--413},
address = {Abu Dhabi, United Arab Emirates},
publisher = {Association for Computational Linguistics},
}

@article{loeffler2024reinvent,
  title = {Reinvent 4: {Modern} {AI}–driven Generative Molecule Design},
  volume = {16},
  doi = {10.1186/s13321-024-00812-5},
  number = {1},
  journal = {Journal of Cheminformatics},
  author = {Loeffler, Hannes H. and He, Jiazhen and Tibo, Alessandro and Janet, Jon Paul and Voronov, Alexey and Mervin, Lewis H. and Engkvist, Ola},
  year = {2024},
  pages = {20}
}

@article{jensen2019graphga,
  author = {Jensen, Jan H.},
  title = {A Graph-based Genetic Algorithm and Generative Model/Monte Carlo Tree Search for the Exploration of Chemical Space},
  journal = {Chemical Science},
  year = {2019},
  volume = {10},
  pages = {3567-3572},
  doi = {10.1039/C8SC05372C}
}

@inproceedings{kim2024gflownet,
  author = {Kim, Hyeonah and Kim, Minsu and Choi, Sanghyeok and Park, Jinkyoo},
  booktitle = {Advances in Neural Information Processing Systems},
  pages = {42618--42648},
  title = {Genetic-guided GFlowNets for Sample Efficient Molecular Optimization},
  year = {2024},
  doi = {10.52202/079017-1350},
  publisher = {Curran Associates, Inc.},
  address   = {Red Hook, NY, USA}
}

@inproceedings{gao2022benchmark,
  author = {Gao, Wenhao and Fu, Tianfan and Sun, Jimeng and Coley, Connor},
  booktitle = {Advances in Neural Information Processing Systems},
  pages = {21342--21357},
  title = {Sample Efficiency Matters: A Benchmark for Practical Molecular Optimization},
  volume = {35},
  year = {2022},
  doi = {10.52202/068431-1551},
  publisher = {Curran Associates, Inc.},
  address = {Red Hook, NY, USA}
}

@article{zhang2024deeplead,
  title = {Deep Lead Optimization: Leveraging Generative AI for Structural Modification},
  volume = {146},
  doi = {10.1021/jacs.4c11686},
  number = {46},
  journal = {Journal of the American Chemical Society},
  author = {Zhang, Odin and Lin, Haitao and Zhang, Hui and Zhao, Huifeng and Huang, Yufei and Hsieh, Chang-Yu and Pan, Peichen and Hou, Tingjun},
  year = {2024},
  pages = {31357--31370}
}

@inproceedings{brown2020language,
  author = {Brown, Tom and Mann, Benjamin and Ryder, Nick and Subbiah, Melanie and Kaplan, Jared D and Dhariwal, Prafulla and Neelakantan, Arvind and Shyam, Pranav and Sastry, Girish and Askell, Amanda and Agarwal, Sandhini and Herbert-Voss, Ariel and Krueger, Gretchen and Henighan, Tom and Child, Rewon and Ramesh, Aditya and Ziegler, Daniel and Wu, Jeffrey and Winter, Clemens and Hesse, Chris and Chen, Mark and Sigler, Eric and Litwin, Mateusz and Gray, Scott and Chess, Benjamin and Clark, Jack and Berner, Christopher and McCandlish, Sam and Radford, Alec and Sutskever, Ilya and Amodei, Dario},
  booktitle = {Advances in Neural Information Processing Systems},
  pages = {1877--1901},
  title = {Language Models are Few-Shot Learners},
  year = {2020},
  url = {https://arxiv.org/abs/2005.14165},
  publisher = {Curran Associates, Inc.},
  address   = {Red Hook, NY, USA}
}

@article{liu2023pretrainpromptpredict,
author = {Liu, Pengfei and Yuan, Weizhe and Fu, Jinlan and Jiang, Zhengbao and Hayashi, Hiroaki and Neubig, Graham},
title = {Pre-train, Prompt, and Predict: A Systematic Survey of Prompting Methods in Natural Language Processing},
year = {2023},
publisher = {Association for Computing Machinery},
address = {New York, NY, USA},
volume = {55},
number = {9},
issn = {0360-0300},
url = {https://doi.org/10.1145/3560815},
doi = {10.1145/3560815},
journal = {ACM Comput. Surv.},
month = jan,
articleno = {195},
numpages = {35},
}

@inproceedings{zamfirescu2023whycantprompt,
author = {Zamfirescu-Pereira, J.D. and Wong, Richmond Y. and Hartmann, Bjoern and Yang, Qian},
title = {Why Johnny Can’t Prompt: How Non-AI Experts Try (and Fail) to Design LLM Prompts},
year = {2023},
isbn = {9781450394215},
publisher = {Association for Computing Machinery},
address = {New York, NY, USA},
doi = {10.1145/3544548.3581388},
booktitle = {Proceedings of the 2023 CHI Conference on Human Factors in Computing Systems},
articleno = {437},
numpages = {21},
}

@inproceedings{wu2022aichains,
  author = {Wu, Tongshuang and Terry, Michael and Cai, Carrie Jun},
  title = {AI Chains: Transparent and Controllable Human-AI Interaction by Chaining Large Language Model Prompts},
  year = {2022},
  doi = {10.1145/3491102.3517582},
  booktitle = {Proceedings of the 2022 CHI Conference on Human Factors in Computing Systems},
  numpages = {22},
  publisher = {Association for Computing Machinery}, address = {New York, NY, USA}
}

@article{zhang2024chemnav,
  title = {ChemNav: An Interactive Visual Tool to Navigate in the Latent Space for Chemical Molecules Discovery},
  journal = {Visual Informatics},
  volume = {8},
  number = {4},
  pages = {60-70},
  year = {2024},
  doi = {10.1016/j.visinf.2024.10.002},
  author = {Yang Zhang and Jie Li and Xu Chao}
}

@article{zheng2023desirable,
  title = {Desirable Molecule Discovery via Generative Latent Space Exploration},
  journal = {Visual Informatics},
  volume = {7},
  number = {4},
  pages = {13-21},
  year = {2023},
  doi = {10.1016/j.visinf.2023.10.002},
  author = {Wanjie Zheng and Jie Li and Yang Zhang}
}

@article{menke2024metis,
  title = {Metis: A Python-based User Interface to Collect Expert Feedback for Generative Chemistry Models},
  volume = {16},
  doi = {10.1186/s13321-024-00892-3},
  number = {1},
  journal = {Journal of Cheminformatics},
  author = {Menke, Janosch and Nahal, Yasmine and Bjerrum, Esben Jannik and Kabeshov, Mikhail and Kaski, Samuel and Engkvist, Ola},
  year = {2024},
  pages = {100}
}

@inproceedings{peng2024designprompt,
author = {Peng, Xiaohan and Koch, Janin and Mackay, Wendy E.},
title = {DesignPrompt: Using Multimodal Interaction for Design Exploration with Generative AI},
year = {2024},
publisher = {Association for Computing Machinery},
address = {New York, NY, USA},
doi = {10.1145/3643834.3661588},
booktitle = {Proceedings of the 2024 ACM Designing Interactive Systems Conference},
pages = {804–818},
numpages = {15}
}

@inproceedings{choi2024creativeconnect,
author = {Choi, DaEun and Hong, Sumin and Park, Jeongeon and Chung, John Joon Young and Kim, Juho},
title = {CreativeConnect: Supporting Reference Recombination for Graphic Design Ideation with Generative AI},
year = {2024},
publisher = {Association for Computing Machinery},
address = {New York, NY, USA},
doi = {10.1145/3613904.3642794},
booktitle = {Proceedings of the 2024 CHI Conference on Human Factors in Computing Systems},
articleno = {1055},
numpages = {25},
series = {CHI '24}
}

@inproceedings{chung2023promptpaint,
author = {Chung, John Joon Young and Adar, Eytan},
title = {PromptPaint: Steering Text-to-Image Generation Through Paint Medium-like Interactions},
year = {2023},
isbn = {9798400701320},
publisher = {Association for Computing Machinery},
address = {New York, NY, USA},
doi = {10.1145/3586183.3606777},
booktitle = {Proceedings of the 36th Annual ACM Symposium on User Interface Software and Technology},
articleno = {6},
numpages = {17},
}

@article{wu2018moleculenet,
  author = {Wu, Zhenqin and Ramsundar, Bharath and Feinberg, Evan N and Gomes, Joseph and Geniesse, Caleb and Pappu, Aneesh and Leswing, Karl and Pande, Vijay},
  title = {MoleculeNet: A Benchmark for Molecular Machine Learning},
  journal = {Chemical Science},
  year = {2018},
  volume = {9},
  pages = {513--530},
  doi = {10.1039/C7SC02664A}
}

@misc{ketcher,
  title = {Ketcher: Open-source Web-based Chemical Structure Editor},
  author = {{EPAM Systems}},
  year = {2024},
  url = {https://github.com/epam/ketcher}
}

@article{beck2022small,
  title = {Small Molecules and Their Impact in Drug Discovery: A Perspective on the Occasion of the 125th Anniversary of the Bayer Chemical Research Laboratory},
  author = {Hartmut Beck and Michael Härter and Bastian Haß and Carsten Schmeck and Lars Baerfacker},
  journal = {Drug Discovery Today},
  volume = {27},
  number = {6},
  pages = {1560-1574},
  year = {2022},
  doi = {10.1016/j.drudis.2022.02.015}
}

@article{maurer2022designing,
  title = {Designing Small Molecules for Therapeutic Success: A Contemporary Perspective},
  author = {Tristan S. Maurer and Martin Edwards and David Hepworth and Patrick Verhoest and Charlotte M.N. Allerton},
  journal = {Drug Discovery Today},
  volume = {27},
  number = {2},
  pages = {538-546},
  year = {2022},
  doi = {10.1016/j.drudis.2021.09.017}
}

@article{racz2025changing,
  title = {The Changing Landscape of Medicinal Chemistry Optimization},
  volume = {24},
  doi = {10.1038/s41573-025-01225-1},
  number = {11},
  journal = {Nature Reviews Drug Discovery},
  author = {Rácz, Anita and Mihalovits, Levente M. and Beckers, Maximilian and Fechner, Nikolas and Stiefl, Nikolaus and Sirockin, Finton and McCoull, William and Evertsson, Emma and Lemurell, Malin and Makara, Gergely and Keserű, György M.},
  year = {2025},
  pages = {870--887}
}

@article{hann2012finding,
  title = {Finding the Sweet Spot: the Role of Nature and Nurture in Medicinal Chemistry},
  volume = {11},
  doi = {10.1038/nrd3701},
  number = {5},
  journal = {Nature Reviews Drug Discovery},
  author = {Hann, Michael M. and Keserü, György M.},
  year = {2012},
  pages = {355--365}
}

@article{bickerton2012quantifying,
  title = {Quantifying the Chemical Beauty of Drugs},
  volume = {4},
  number = {2},
  journal = {Nature Chemistry},
  author = {Bickerton, G. Richard and Paolini, Gaia V. and Besnard, Jérémy and Muresan, Sorel and Hopkins, Andrew L.},
  year = {2012},
  pages = {90--98},
  doi = {10.1038/nchem.1243}
}

@article{bran2024augmenting,
  title = {Augmenting Large Language Models with Chemistry Tools},
  author = {Andres M Bran and Sam Cox and Oliver Schilter and Carlo Baldassari and Andrew D White and Philippe Schwaller},
  journal = {Nature Machine Intelligence},
  year = {2024},
  volume = {6},
  number = {5},
  pages = {525--535},
  doi = {10.1038/s42256-024-00832-8}
}

@article{schwaller2021prediction,
  title = {Mapping the Space of Chemical Reactions using Attention-based Neural Networks},
  volume = {3},
  doi = {10.1038/s42256-020-00284-w},
  number = {2},
  journal = {Nature Machine Intelligence},
  author = {Schwaller, Philippe and Probst, Daniel and Vaucher, Alain C. and Nair, Vishnu H. and Kreutter, David and Laino, Teodoro and Reymond, Jean-Louis},
  year = {2021},
  pages = {144--152}
}

@article{ozcelik2025generative,
  author = {\"{O}z{\c{c}}elik, Riza and Brinkmann, Helena and Criscuolo, Emanuele and Grisoni, Francesca},
  title = {Generative Deep Learning for de Novo Drug Design-A Chemical Space Odyssey},
  year = {2025},
  volume = {65},
  number = {14},
  doi = {10.1021/acs.jcim.5c00641},
  journal = {Journal of Chemical Information and Modeling},
  pages = {7352--7372}
}

@article{stokes2020deep,
  title = {A Deep Learning Approach to Antibiotic Discovery},
  volume = {180},
  doi = {10.1016/j.cell.2020.01.021},
  number = {4},
  journal = {Cell},
  author = {Stokes, Jonathan M. and Yang, Kevin and Swanson, Kyle and Jin, Wengong and Cubillos-Ruiz, Andres and Donghia, Nina M. and MacNair, Craig R. and French, Shawn and Carfrae, Lindsey A. and Bloom-Ackermann, Zohar and Tran, Victoria M. and Chiappino-Pepe, Anush and Badran, Ahmed H. and Andrews, Ian W. and Chory, Emma J. and Church, George M. and Brown, Eric D. and Jaakkola, Tommi S. and Barzilay, Regina and Collins, James J.},
  year = {2020},
  pages = {688--702.e13}
}

@inproceedings{arawjo2024chainforge,
  author    = {Arawjo, Ian and Swoopes, Chelse and Vaithilingam, Priyan and Wattenberg, Martin and Glassman, Elena L.},
  title     = {{ChainForge}: A Visual Toolkit for Prompt Engineering and {LLM} Hypothesis Testing},
  booktitle = {Proceedings of the 2024 CHI Conference on Human Factors in Computing Systems},
  year      = {2024},
  publisher = {Association for Computing Machinery},
  address   = {New York, NY, USA},
  isbn      = {9798400703300},
  doi       = {10.1145/3613904.3642016},
  url       = {https://doi.org/10.1145/3613904.3642016},
  articleno = {304},
  numpages  = {18},
  location  = {Honolulu, HI, USA},
  series    = {CHI '24}
}

@article{Zhu2024promptbench,
  author  = {Kaijie Zhu and Qinlin Zhao and Hao Chen and Jindong Wang and Xing Xie},
  title   = {PromptBench: A Unified Library for Evaluation of Large Language Models},
  year    = {2024},
  volume  = {25},
  number  = {254},
  pages   = {1--22},
  journal = {Journal of Machine Learning Research},
  url={https://arxiv.org/abs/2312.07910}
}

@article{jiang2022prompting,
author = {Beurer-Kellner, Luca and Fischer, Marc and Vechev, Martin},
title = {Prompting Is Programming: A Query Language for Large Language Models},
year = {2023},
issue_date = {June 2023},
publisher = {Association for Computing Machinery},
address = {New York, NY, USA},
volume = {7},
number = {PLDI},
url = {https://doi.org/10.1145/3591300},
doi = {10.1145/3591300},
journal = {Proc. ACM Program. Lang.},
month = jun,
articleno = {186},
numpages = {24}
}

@article{tabana2023target,
  title = {Target Identification of Small Molecules: An Overview of the Current Applications in Drug Discovery},
  volume = {23},
  number = {1},
  journal = {BMC Biotechnology},
  author = {Tabana, Yasser and Babu, Dinesh and Fahlman, Richard and Siraki, Arno G. and Barakat, Khaled},
  year = {2023},
  pages = {44},
  doi = {10.1186/s12896-023-00815-4}
}

@article{daina2017swissadme,
  author = {Daina, Antoine and Michielin, Olivier and Zoete, Vincent},
  title = {{SwissADME: a free web tool to evaluate pharmacokinetics, drug-likeness and medicinal chemistry friendliness of small molecules}},
  journal = {Scientific Reports},
  year = {2017},
  volume = {7},
  number = {1},
  pages = {42717},
  doi = {10.1038/srep42717}
}

@article{xiong2021admetlab,
  title = {ADMETlab 2.0: An Integrated Online Platform for Accurate and Comprehensive Predictions of ADMET Properties},
  author = {Xiong, Guoli and Wu, Zhenxing and Yi, Jiacai and Fu, Li and Yang, Zhijiang and Hsieh, Changyu and Yin, Mingzhu and Zeng, Xiangxiang and Wu, Chengkun and Lu, Aiping and others},
  journal = {Nucleic acids research},
  volume = {49},
  number = {W1},
  pages = {W5--W14},
  year = {2021},
  doi = {10.1093/nar/gkab255}
}

@incollection{brooke1996sus,
  author    = {John Brooke},
  title     = {SUS: A ``Quick and Dirty'' Usability Scale},
  booktitle = {Usability Evaluation in Industry},
  editor    = {Patrick W. Jordan and Bruce Thomas and Bernard A. Weerdmeester and Ian L. McClelland},
  pages     = {189--194},
  year      = {1996},
  publisher = {Taylor \& Francis},
  address   = {London}
}

@article{lipinski2012experimental,
  title = {Experimental and Computational Approaches to Estimate Solubility and Permeability in Drug Discovery and Development Settings},
  author = {Lipinski, Christopher A and Lombardo, Franco and Dominy, Beryl W and Feeney, Paul J},
  journal = {Advanced drug delivery reviews},
  volume = {64},
  pages = {4--17},
  year = {2012},
  doi = {10.1016/j.addr.2012.09.019}
}

@article{veber2002molecular,
  title = {Molecular Properties that Influence the Oral Bioavailability of Drug Candidates},
  author = {Veber, Daniel F and Johnson, Stephen R and Cheng, Hung-Yuan and Smith, Brian R and Ward, Keith W and Kopple, Kenneth D},
  journal = {Journal of medicinal chemistry},
  volume = {45},
  number = {12},
  pages = {2615--2623},
  year = {2002},
  doi = {10.1021/jm020017n}
}

@misc{rdkit,
  author       = {Landrum, Greg and others},
  title        = {{RDKit}: Open-Source Cheminformatics Software},
  year         = {2026},
  howpublished = {Zenodo},
  doi          = {10.5281/zenodo.21291217}
}

@article{shrout1979intraclass,
  title = {Intraclass correlations: uses in assessing rater reliability},
  author = {Shrout, Patrick E. and Fleiss, Joseph L.},
  journal = {Psychological Bulletin},
  volume = {86},
  number = {2},
  pages = {420--428},
  year = {1979},
  doi = {10.1037/0033-2909.86.2.420}
}

@inproceedings{masson2024directgpt,
author = {Masson, Damien and Malacria, Sylvain and Casiez, G\'{e}ry and Vogel, Daniel},
title = {DirectGPT: A Direct Manipulation Interface to Interact with Large Language Models},
year = {2024},
publisher = {Association for Computing Machinery},
address = {New York, NY, USA},
doi = {10.1145/3613904.3642462},
booktitle = {Proceedings of the 2024 CHI Conference on Human Factors in Computing Systems},
articleno = {975},
numpages = {16}
}

@article{ishida2025chatchemts,
  author = {Ishida, Shoichi and Sato, Tomohiro and Honma, Teruki and Terayama, Kei},
  title = {Large Language Models Open New Way of {AI}-Assisted Molecule Design for Chemists},
  journal = {Journal of Cheminformatics},
  volume = {17},
  number = {1},
  pages = {36},
  year = {2025},
  doi = {10.1186/s13321-025-00984-8}
}

@inproceedings{kim2023cells,
  author = {Kim, Tae Soo and Lee, Yoonjoo and Chang, Minsuk and Kim, Juho},
  title = {Cells, Generators, and Lenses: Design Framework for Object-Oriented Interaction with Large Language Models},
  booktitle = {Proceedings of the 36th Annual ACM Symposium on User Interface Software and Technology},
  articleno = {4},
  numpages = {18},
  year = {2023},
  doi = {10.1145/3586183.3606833},
  publisher = {Association for Computing Machinery}, address = {New York, NY, USA},
}

@inproceedings{sun2025kerag,
  author    = {Sun, Yushi and Sun, Kai and Xu, Yifan Ethan and Yang, Xiao and Dong, Xin Luna and Tang, Nan and Chen, Lei},
  title     = {{KERAG}: Knowledge-Enhanced Retrieval-Augmented Generation for Advanced Question Answering},
  booktitle = {Findings of the Association for Computational Linguistics: EMNLP 2025},
  year      = {2025},
  publisher = {Association for Computational Linguistics},
  address   = {Suzhou, China},
  pages     = {6194--6216},
  doi       = {10.18653/v1/2025.findings-emnlp.329}
}

@inproceedings{yang2024crag,
 author = {Yang, Xiao and Sun, Kai and Xin, Hao and Sun, Yushi and Bhalla, Nikita and Chen, Xiangsen and Choudhary, Sajal and Gui, Rongze Daniel and Jiang, Ziran Will and Jiang, Ziyu and Kong, Lingkun and Moran, Brian and Wang, Jiaqi and Xu, Yifan Ethan and Yan, An and Yang, Chenyu and Yuan, Eting and Zha, Hanwen and Tang, Nan and Chen, Lei and Scheffer, Nicolas and Liu, Yue and Shah, Nirav and Wanga, Rakesh and Kumar, Anuj and Yih, Wen-tau and Dong, Xin Luna},
 booktitle = {Advances in Neural Information Processing Systems},
 doi = {10.52202/079017-0335},
 pages = {10470--10490},
 publisher = {Curran Associates, Inc.},
 address   = {Red Hook, NY, USA},
 title = {CRAG - Comprehensive RAG Benchmark},
 volume = {37},
 year = {2024}
}

@inproceedings{sheng2023knowledge,
author = {Sheng, Rui and Yang, Leni and Li, Haotian and Luo, Yan and Xu, Ziyang and Zhou, Zhilan and Gotz, David and Qu, Huamin},
title = {Knowledge Compass: A Question Answering System Guiding Students with Follow-Up Question Recommendations},
year = {2023},
isbn = {9798400700965},
publisher = {Association for Computing Machinery},
address = {New York, NY, USA},
url = {https://doi.org/10.1145/3586182.3615785},
doi = {10.1145/3586182.3615785},
booktitle = {Adjunct Proceedings of the 36th Annual ACM Symposium on User Interface Software and Technology},
articleno = {65},
numpages = {4},
location = {San Francisco, CA, USA},
series = {UIST '23 Adjunct}
}

@article{sheng2026designpatterns,
title = {Design patterns of human-AI interfaces in healthcare},
journal = {International Journal of Human-Computer Studies},
volume = {209},
pages = {103737},
year = {2026},
issn = {1071-5819},
doi = {https://doi.org/10.1016/j.ijhcs.2026.103737},
url = {https://www.sciencedirect.com/science/article/pii/S1071581926000121},
author = {Rui Sheng and Chuhan Shi and Sobhan Lotfi and Shiyi Liu and Adam Perer and Huamin Qu and Furui Cheng}
}

@ARTICLE{sheng2025trialcompass,
  author={Sheng, Rui and Wang, Xingbo and Wang, Jiachen and Jin, Xiaofu and Sheng, Zhonghua and Xu, Zhenxing and Rajendran, Suraj and Qu, Huamin and Wang, Fei},
  journal={IEEE Transactions on Visualization and Computer Graphics}, 
  title={TrialCompass: Visual Analytics for Enhancing the Eligibility Criteria Design of Clinical Trials}, 
  year={2026},
  volume={32},
  number={1},
  pages={1230-1240},
  doi={10.1109/TVCG.2025.3634803}}
\end{document}